\documentclass[%
 reprint,
superscriptaddress,
 amsmath,amssymb,
 aps,
]{revtex4-2}

\usepackage{graphicx}
\usepackage{dcolumn}
\usepackage{bm}
\usepackage{extpfeil}
\usepackage{comment}
\usepackage{xcolor}
\usepackage{hyperref}

\newcommand{\dee}{\mathrm{d}}
\newcommand{\ba}{\begin{eqnarray}}
\newcommand{\ea}{\end{eqnarray}}

\begin{document}
\title{Statistics of thermal gas pressure as a probe of cosmology and galaxy formation} 
\author{Ziyang Chen}
\email{chen\_zy@sjtu.edu.cn}
\affiliation{%
 Department of Astronomy, School of Physics and Astronomy, Shanghai Jiao Tong University, Shanghai, 200240, China
}%
\affiliation{%
Key Laboratory for Particle Astrophysics and Cosmology (MOE)/Shanghai Key Laboratory for Particle Physics and Cosmology, China
}%
\affiliation{%
Max Planck Institute f\"ur Astrophysik, Karl-Schwarzschild-Str. 1, 85748 Garching, Germany
}%
\author{Drew Jamieson}%
\email{jamieson@mpa-garching.mpg.de}
\affiliation{%
Max Planck Institute f\"ur Astrophysik, Karl-Schwarzschild-Str. 1, 85748 Garching, Germany
}%
\author{Eiichiro Komatsu}
\affiliation{%
Max Planck Institute f\"ur Astrophysik, Karl-Schwarzschild-Str. 1, 85748 Garching, Germany
}%
\affiliation{%
Ludwig-Maximilians-Universit\"at M\"unchen, Schellingstr. 4, 80799 M\"unchen, Germany
}
\affiliation{%
Kavli IPMU (WPI), UTIAS, The University of Tokyo, Kashiwa, 277-8583, Japan
}
\author{Sownak Bose}
\affiliation{%
Institute for Computational Cosmology, Department of Physics, Durham University, South Road, Durham, DH1 3LE, UK
}%
\author{Klaus Dolag}
\affiliation{%
Universit\"ats-Sternwarte, Fakult\"at f\"ur Physik, Ludwig-Maximilians-Universit\"at M\"unchen, Scheinerstr. 1, 81679 M\"unchen, Germany
}%
\affiliation{%
Max Planck Institute f\"ur Astrophysik, Karl-Schwarzschild-Str. 1, 85748 Garching, Germany
}%
\author{Boryana Hadzhiyska}
\affiliation{%
Lawrence Berkeley National Laboratory, 1 Cyclotron Road, Berkeley, CA 94720, USA
}%
\affiliation{%
University of California, Berkeley, 110 Sproul Hall \#5800 Berkeley, CA 94720, USA
}%
\author{C\'esar Hern\'andez-Aguayo}
\affiliation{%
Max Planck Institute f\"ur Astrophysik, Karl-Schwarzschild-Str. 1, 85748 Garching, Germany
}%
\affiliation{%
Excellence Cluster ORIGINS, Boltzmannstr. 2, 85748 Garching, Germany
}%
\author{Lars Hernquist}
\affiliation{%
Harvard-Smithsonian Center for Astrophysics, 60 Garden Street, Cambridge, MA 02138, USA
}%
\author{Rahul Kannan}
\affiliation{%
Department of Physics and Astronomy, York University, 4700 Keele Street, Toronto, ON M3J 1P3, Canada
}%
\author{R\"udiger Pakmor}
\affiliation{%
Max Planck Institute f\"ur Astrophysik, Karl-Schwarzschild-Str. 1, 85748 Garching, Germany
}%
\author{Volker Springel}
\affiliation{%
Max Planck Institute f\"ur Astrophysik, Karl-Schwarzschild-Str. 1, 85748 Garching, Germany
}%
\affiliation{%
Ludwig-Maximilians-Universit\"at M\"unchen, Schellingstr. 4, 80799 M\"unchen, Germany
}

\date{\today}

\begin{abstract}
The statistics of thermal gas pressure are a new and promising probe of cosmology and astrophysics. The large-scale cross-correlation between galaxies and the thermal Sunyaev-Zeldovich effect gives the bias-weighted mean electron pressure, $\langle b_\mathrm{h}P_e\rangle$. In this paper, we show that $\langle b_\mathrm{h}P_e\rangle$ is sensitive to the amplitude of fluctuations in matter density, for example $\langle b_\mathrm{h}P_e\rangle\propto \left(\sigma_8\Omega_\mathrm{m}^{0.81}h^{0.67}\right)^{3.14}$ at redshift $z=0$. 
We find that at $z<0.5$ the observed $\langle b_\mathrm{h}P_e\rangle$ is smaller than that predicted by the state-of-the-art hydrodynamical simulations of galaxy formation, MillenniumTNG, by a factor of $0.93$. 
This can be explained by a lower value of $\sigma_8$ and $\Omega_\mathrm{m}$, similar to the so-called ``$S_8$ tension'' seen in the gravitational lensing effect, although the influence of astrophysics cannot be completely excluded. The difference between \emph{Magneticum} and MillenniumTNG at $z<2$ is small, indicating that the difference in the galaxy formation models used by these simulations has little impact on $\langle b_\mathrm{h}P_e\rangle$ at this redshift range. At higher $z$, we find that both simulations are in a modest tension with the existing upper bounds on $\langle b_\mathrm{h}P_e\rangle$. We also find a significant difference between these simulations there, which we attribute to a larger sensitivity to the galaxy formation models in the high redshift regime. Therefore, more precise measurements of $\langle b_\mathrm{h}P_e\rangle$ at all redshifts will provide a new test of our understanding of cosmology and galaxy formation.
\end{abstract}

                             
\maketitle

\section{Introduction}\label{sec1}

    The traditional tools of cosmological inference during the era of late-time large-scale structure (LSS) are concerned primarily with the background expansion history and growth history of the inhomogeneities that comprise the cosmic web. By constructing the Hubble-Lema\^{\i}tre diagram, observing the scale of the Baryon Acoustic Oscillations (BAO) peak \cite{eBOSS:2021}, we obtain ever tighter constraints on the background parameters governing the Universe's expansion. From matter clustering statistics, such as the two-point statistics obtained from weak gravitational lensing measurements \cite{DES:2022,HSC2023,KiDS:2023}, we obtain further constraints on the amplitude of fluctuations in the distribution of matter. 
    
    These techniques have achieved considerable success in extracting precision measurements of the cosmological parameters, which in turn has revealed tension with cosmic microwave background (CMB) observations for both the Hubble parameter, $H_0=100~h$~km/s/Mpc \cite{Abdalla2022}, and the amplitude of density fluctuations, $S_8$ \cite{Dragan2023}. One strategy toward resolving these tensions is to broaden the scope of cosmological observables used for inference. In addition to the expansion and growth histories, we can consider the thermal history of the Universe \cite{Chiang2020,Chiang2021,Young2021}.

    The thermal Sunyaev-Zeldovich (tSZ) effect \cite{SZ1972} provides a powerful tool for probing the thermal properties of the Universe. CMB photons inverse-Compton scatter off of hot gas in massive galaxy clusters, creating secondary anisotropies that can be extracted from CMB maps. The resulting Compton-$y$ map provides compressed information, integrated along the line of sight, for the distribution of electron pressure in an ionized gas. By cross-correlating the Compton-$y$ map with other LSS tracers such as galaxies, galaxy groups, Lyman-$\alpha$ forest, or fast radio bursts, it is possible to disentangle its contributions from different redshifts \cite{Menard2013}. Recent studies \cite{Vikram2017,Pandey2019,Koukoufilippas2020,Chiang2020,Chen2022,Sanchez2022} have utilized this approach to measure the cross-correlation between tSZ and galaxies/groups from observational data. 

    The central challenges of late-time LSS analysis are contending with the nonlinearity of gravitational clustering and the complicated baryonic processes involved in galaxy formation. In the next few decades, several surveys, such as LSST \cite{LSST2019}, Euclid \cite{Euclid2011}, Roman Space Telescope \cite{WFIRST2013}, CSST \cite{CSST2023}, DESI \cite{DESI2016} and PFS \cite{Takada2014}, are expected to constrain the cosmological models to an accuracy approaching 1\% \cite{vanDaalen2011}. The development of accurate galaxy formation and evolution models is crucial to minimize theoretical systematic errors and ensure that they are smaller than the measurement errors. Since the tSZ signal arises from ionized gas in massive halos, it is particularly sensitive to this physics. For example, Ref.~\cite{Hadzhiyska2023} finds that the $M-Y$ relation, where $Y$ is the integrated Compton-$y$ parameter, is sensitive to models of baryonic feedback.

    Hydrodynamical simulations are now widely used to study galaxy formation and evolution in a cosmological context \cite{Illustris,EAGLE, Magneticum,McCarthy2017,ITNG,SIMBA,Pakmor:2023,ASTRID,FLAMINGO}. To keep a balance between the computational cost and the accuracy of modeling baryonic processes, many simulations utilize subgrid recipes for phenomena such as stellar winds, Active Galactic Nuclei (AGN) feedback, and star formation \cite{Somerville2015}. Although these recipes vary between simulation codes, they are all calibrated to reproduce basic galaxy observations in some way. Discriminating between different galaxy formation models remains a major area of interest.

    In this paper, we demonstrate that tSZ$\times$LSS observables are capable of distinguishing between different implementations of baryonic feedback processes in the state-of-art of cosmological hydrodynamical simulations. We analyze the clustering statistics of electron pressure in two hydrodynamical simulations that incorporate different galaxy formation models, MillenniumTNG \cite{Pakmor:2023,Hernandez-Aguayo:2023} and \emph{Magneticum} \cite{Magneticum}, and find a significant discrepancy at high redshift. This demonstrates that future tSZ$\times$LSS observations can discriminate between and rule out galaxy formation models.

    We also find that the low redshift clustering statistics of electron pressure from the two simulations agree with each other, and both are slightly, but systematically, higher than the observations. This is similar to, and may be related to, the $S_8$ tension from weak lensing \cite{Dragan2023}. Using halo model calculations, we explore the cosmology and redshift dependence of this tSZ observable. We find that information from the clustering of electron pressure at various redshifts can break the parameter degeneracies between $\sigma_8$, $\Omega_\mathrm{m}$, and $h$ at a single redshift. This demonstrates that tSZ$\times$LSS observations could be used as a new probe of cosmology. In particular, future measurements of the thermal history of the Universe may provide new information about the $S_8$ tension.

    The outline of this paper is as follows. In Sec.~\ref{sec2}, we discuss pressure statistics and current constraints from observations. In Sec~\ref{sec3}, we introduce the simulations adopted in the work and the measurement of $\langle b_\mathrm{h}P_e \rangle$ in them.  We then study the reasons behind the behavior of $\langle b_\mathrm{h}P_e \rangle$ in Sec.~\ref{sec4}. In Sec~\ref{sec5}, we list caveats and discuss high-redshift measuements of $\langle b_\mathrm{h}P_e \rangle$. We conclude in Sec.~\ref{sec6}.

\section{Pressure statistics}\label{sec2}
    
    When CMB photons scatter off of electrons in an ionized gas, their temperature fluctuations are distorted. The secondary temperature fluctuations induced by the tSZ effect can be expressed as \cite{Mroczkowski2019}
    \ba
    \frac{\Delta T(\hat{n}, \nu)}{T_{\rm CMB}} = y(\hat{n}) f(x)  \, ,
    \label{eq:tsz_v1}
    \ea
    where $\hat n$ is the line-of-sight direction vector and 
    \ba
    x=\frac{h\nu}{k_\mathrm{B}T_{\rm CMB}} \, .
    \label{eq:tsz_v2}
    \ea
    Here, $h$ is the Planck constant (not to be confused with the reduced Hubble constant); $\nu$ is the frequency of photons; $k_{\rm B}$ is the Boltzmann constant; and $T_{\rm CMB}=2.725~\rm K $ is the CMB temperature \cite{Fixsen2009}. Assuming that the gas is non-relativistic, the frequency dependence is \cite{SZ1969}
    \ba
    f(x) = x\frac{e^x+1}{e^x-1}-4 \, .
    \ea
    This distinctive frequency dependence allows the tSZ component to be separated from the CMB maps using multi-frequency measurements~\cite{Planck2016_tSZ}.
    
    The $y$ parameter is found by integrating the electron pressure along the line of sight
    \ba
    y(\hat n) = \frac{\sigma_\mathrm{T}}{m_ec^2}\int_0^{z_{\mathrm{re}}} \frac{\dee z}{1+z} \frac{\dee\chi}{\dee z} P_e\left[ \chi(z) \hat{n}, z \right] \, , 
    \label{eq:cy}
    \ea
    where $\sigma_\mathrm{T}$ is the Thomson scattering cross section; $m_e$ is the electron mass; $c$ is the light speed; $\chi(z)$ is the co-moving distance to redshift $z$; and $P_e = k_\mathrm{B}n_eT_e$ is the electron pressure. The pressure is integrated between now and $z_{\mathrm{re}}\simeq 1090$, the redshift of recombination.

    \subsection{Pressure statistics at a fixed redshift}

    At a fixed redshift, the spatial distribution of pressure is described by the field $P_e(\mathbf{x}, z)$, where $\mathbf{x}$ are comoving coordinates. We decompose this field into inhomogeneities, $\delta P_{e}(\mathbf{x}, z)$, fluctuating around a redshift-dependent mean, $\bar{P}_{e}(z)$,
    \begin{align}
        P_e(\mathbf{x}, z) = \bar{P}_{e}(z) + \delta P_{e}(\mathbf{x}, z) \, .
    \end{align}
    We Fourier transform the fluctuations to obtain their modes $\delta P_e(\mathbf{k}, z)$, where $\mathbf{k}$ is the comoving wave vector. We use the same symbol to denote a field in both coordinate and Fourier space, distinguishing them with their arguments.

    The matter density field is 
    \ba
    \rho_{\mathrm{m}}(\mathbf{x}, z) = \bar{\rho}_\mathrm{m}(z)\left[ 1 + \delta_{\mathrm{m}}(\mathbf{x}, z)\right] \, ,
    \ea
    where $\bar{\rho}_\mathrm{m}(z)$ is the mean matter density and $\delta_{\mathrm{m}}(\mathbf{x}, z)$ is the density contrast, which has modes $\delta_{\mathrm{m}}(\mathbf{k}, z)$. Electron pressure is a biased tracer of the matter field, which can be accurately described on large scales by the linear bias expansion \cite{Refregier2000},
    \ba
        \delta P_e(\mathbf{k}, z) \simeq \bar{P}_e(z) b_y(z)  \delta_{\mathrm{m}}(\mathbf{k}, z) \, .
        \label{eq:Pelin}
    \ea
    Here, $b_y(z)$ is the linear bias that we associate with the Compton-$y$ parameter given in Eq.~\eqref{eq:cy}, and $\bar{P}_e(z)$ is the mean electron pressure. For reasons that will become clear below, we define the mean electron pressure weighted by halo bias as
    \ba
        \langle b_\mathrm{h} P_e \rangle(z) = b_y(z) \bar{P}_e(z) \, .
    \ea
    On large scales the electron pressure $\times$ matter cross power spectrum is given by
    \begin{align}{}
        \langle P_e(\mathbf{k}, z) \delta_{\mathrm{m}}(\mathbf{k}', z) \rangle &= (2\pi)^3 \delta_{\rm D}^{(3)}(\mathbf{k} + \mathbf{k}') P_{P_e \mathrm{m}}(k, z), \\ 
         P_{P_e \mathrm{m}}(k, z) &\simeq \langle b_\mathrm{h} P_e \rangle(z) P_{\mathrm{mm}}(k, z) \, .
    \end{align}
    Here, $\delta_{\rm D}^{(3)}(\mathbf{k})$ is Dirac's delta function and $P_{\mathrm{mm}}(k, z)$ is the matter power spectrum. This allows us to relate $\langle b_\mathrm{h} P_e \rangle$ to the large-scale ratio of $P_{P_e \mathrm{m}}$ and $ P_{\mathrm{mm}}$,
    \begin{align}
        \langle b_\mathrm{h} P_e \rangle(z) = \lim_{k\rightarrow0} \frac{P_{P_e \mathrm{m}}(k,z)}{P_{\mathrm{mm}}(k,z)} \, .
        \label{eq:bysim}
    \end{align}
    This expression is the focus of our simulation analysis \cite{Young2021}, since we have direct access to both the electron pressure and the matter field from the simulation snapshots.

    To compare with observations, we consider a galaxy catalog at the same redshift and construct the galaxy number density field, $n_{\mathrm{g}}(\mathbf{x}, z) = \bar{n}_{\mathrm{g}}(z)\left[ 1 + \delta_{\mathrm{g}}(\mathbf{x}, z)\right]$, with the mean number density, $\bar{n}_{\mathrm{g}}(z)$, and the density contrast, $\delta_{\mathrm{g}}(\mathbf{x}, z)$. The modes of the galaxy number density field are well approximated on large scales in terms of the linear galaxy bias $b_{\mathrm{g}}(z)$ \cite{Kaiser1987},
    \begin{align}
        \delta_{\mathrm{g}}(\mathbf{k}, z) \simeq b_{\mathrm{g}}(z) \delta_{\mathrm{m}}(\mathbf{k},z) \, .
    \end{align}
    Then we find the large-scale electron pressure $\times$ galaxy cross power spectrum,
    \begin{align}
       \langle P_e(\mathbf{k}, z) \delta_{\mathrm{g}}(\mathbf{k}', z) \rangle &= (2\pi)^3 \delta_{\rm D}^{(3)}(\mathbf{k} + \mathbf{k}') P_{P_e \mathrm{g}}(k, z), \\ 
         P_{P_e \mathrm{g}}(k, z) &\simeq \langle b_\mathrm{h} P_e \rangle(z) b_{\mathrm{g}}(z) P_{\mathrm{mm}}(k, z) \, .
         \label{eq:hPbg}
    \end{align}

    \subsection{Pressure statistics on the lightcone}

    Now consider a galaxy sample spanning the redshift range $z_1 \leq z \leq z_2$ that overlaps with the map $y(\hat{n})$ in the sky. For this galaxy sample we construct the number density contrast, $\delta_{\mathrm{g}}(\hat{n}, z)$. The Compton-$y$ $\times$ galaxy angular cross power spectrum is
    \begin{align}
   	C_{\ell, y\mathrm{g}}(z_1, z_2) \simeq \int_{z_1}^{z_2} \dee z & 
    \displaystyle\frac{\dee\chi}{\dee z} \displaystyle\frac{W_{y}(z)W_{\mathrm{g}}(z)}{\chi^2(z)} \nonumber \\ & \times P_{P_e\mathrm{g}}\left[(\ell+1/2)\chi^{-1}(z), z\right] \, ,
    \end{align}
    where we have taken $k\simeq(\ell + 1/2)\chi^{-1}$ according to the Limber approximation \cite{LoVerde2008}. The window functions, $W_y$ and $W_\mathrm{g}$, are given by
    \ba
        W_y(z) &=& \frac{\sigma_\mathrm{T}}{m_ec^2}\frac 1 {1+z} \, , \\
        W_\mathrm{g}(z) &=& \frac{H(z)}{c}\phi_\mathrm{g}(z) \, ,
    \ea
    for a galaxy selection function, $\phi_\mathrm{g}(z)$. From Eq.~\eqref{eq:hPbg} we find
    \begin{align}
   	C_{\ell, y\mathrm{g}}(z_1,z_2) \simeq \int_{z_1}^{z_2} \dee z & \displaystyle\frac{\dee\chi}{\dee z} \displaystyle\frac{W_{y}(z)W_{\mathrm{g}}(z)}{\chi^2(z)} b_{\mathrm{g}}(z) \langle b_\mathrm{h} P_e \rangle(z) \nonumber \\ & \times P_{\mathrm{mm}}\left[(\ell+1/2)\chi^{-1}(z), z\right] \, .
    \label{eq:Cygint}
    \end{align}

    \subsection{Pressure statistics from the halo model}

    While Eqs.~(\ref{eq:bysim}) and (\ref{eq:Cygint}) are sufficient for comparing simulations to observations, it is useful to have a physical interpretation of $\langle b_\mathrm{h} P_e \rangle$ as well as a phenomenological framework to predict its value. To this end, we invoke the halo model to describe the distribution of electron pressure at a fixed redshift.

    The simplest form of the halo model (see Ref.~\cite{Cooray2002,Asgari2023} and the references therein for further details) assumes that all matter can be associated with halos through their radial density profiles, $u_{\mathrm{m,h}}(r, z\,|\,M)$,
    \begin{align}
        \frac{\rho_{\mathrm{m}}(\mathbf{x}, z)}{\bar{\rho}_{\mathrm{m}}(z)} = \sum_i u_{\mathrm{m,h}}(|\mathbf{x} - \mathbf{x}_i|, z\,|\,M_i) \, .
    \end{align}
    The sum is over all halos. We denote the Fourier transform of the density profile by $u_{\mathrm{m,h}}(k,z\,|\,M)$, so the modes of the density field are
    \begin{align}
        \frac{\rho_{\mathrm{m}}(\mathbf{k}, z)}{\bar{\rho}_{\mathrm{m}}(z)} = \sum_i e^{-i\mathbf{k}\cdot\mathbf{x}_i} u_{\mathrm{m,h}}(k, z\,|\,M_i) \, .
    \end{align}
    We make analogous assumptions for the electron pressure, introducing the halo pressure profile $u_{P_e,\mathrm{h}}(r, z\,|\,M)$,
    \begin{align}
        \frac{P_{e}(\mathbf{k}, z)}{\bar{P}_{e}(z)} = \sum_i e^{-i\mathbf{k}\cdot\mathbf{x}_i} u_{P_e,\mathrm{h}}(k, z\,|\,M_i) \, .
    \end{align}
    
        \begin{figure*}
        \includegraphics[width = 1\textwidth]{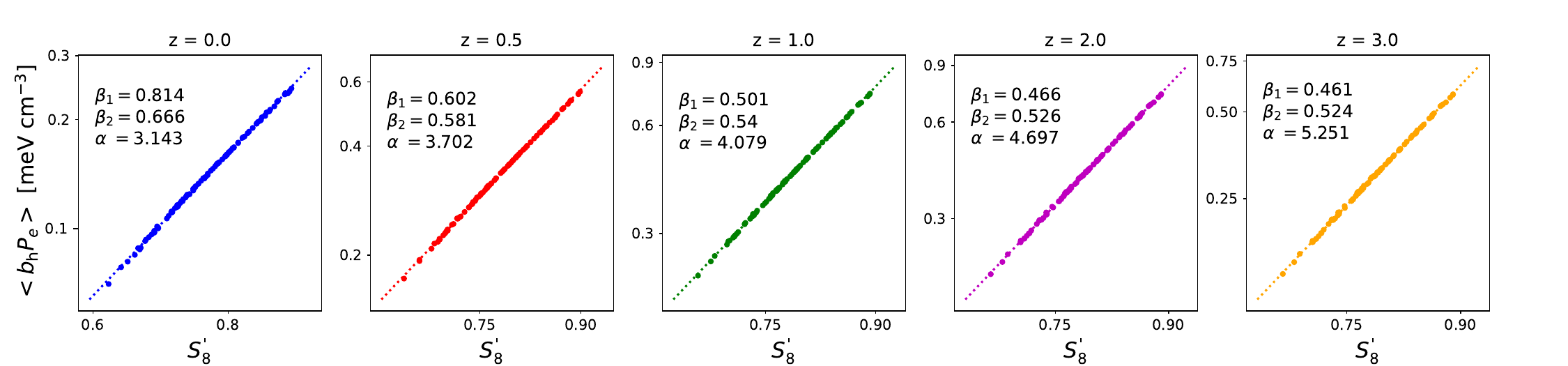}
        \caption{The relation between $\langle b_\mathrm{h}P_e \rangle$ from the halo model [Eq.~\eqref{eq:hmbP}] and the extended $S_8$ parameter defined in Eq.~\eqref{eq:S8prime} at $z=0, 0.5, 1, 2$, and $3$. The points are $\langle b_\mathrm{h}P_e \rangle_{\rm HM}$ with different values of the cosmological parameters. The dashed lines are the results of the fit. The best-fitting values of $\beta_1$, $\beta_2$, and $\alpha $ are shown in the top-left corner of each panel.\label{fig:bPe_S8}}
    \end{figure*}
    
    The cross power spectrum between the electron pressure field and the matter density field has two contributions: a two-halo term correlating the matter from one halo with the pressure from another, and a one-halo term correlating the matter and pressure in the same halo. As described in Appendix \ref{app:halo_model}, this gives
    \ba
        P_{P_e\mathrm{m}}(k, z) & = &\langle b_\mathrm{h}P_e\rangle_{\rm HM}(z) P_{\mathrm{mm}}(k, z)  \nonumber \\
        & +& \int \dee\!\log\!M \frac{\dee n_{\mathrm{h}}}{\dee\!\log\!M}\, 4\pi\! \int r^2 \dee r\, \big[ u_{P_e,\mathrm{h}}(r, z\,|\,M) \nonumber \\
        & &\hspace{80pt} \times u_{\mathrm{m,h}}(r, z\,|\,M)
        \big]\,,
    \ea
    where the subscript ``HM'' denotes the halo model.
    In the one-halo term (the second term in the above expression), $\dee n_{\mathrm{h}}/\dee\!\log\!M$ is the halo mass function~\cite{Tinker2008}. The linear bias of the two-halo term (the first term) is given by \cite{Vikram2017}
    \begin{align}
    \langle b_{\mathrm{h}} P_e \rangle_{\rm HM}(z) =  \bar{P}_e(z) \int & \dee\!\log\!M \frac{\dee n_{\mathrm{h}}}{\dee\!\log\!M} b_\mathrm{h}(M) \nonumber \\ & \times 4\pi \int r^2 \dee r\,  u_{P_e,\mathrm{h}}(r,z\,|\,M) \, ,
    \label{eq:hmbP}
    \end{align}
    where $b_{\mathrm{h}}$ is the linear halo bias \cite{Tinker2010}. We see that in the halo model $\langle b_{\mathrm{h}} P_e \rangle(z)$ can be interpreted as the halo-bias-weighted mean electron pressure.

    Using the halo model described above, with the halo mass function of Ref.~\cite{Tinker2008}, the halo bias of Ref.~\cite{Tinker2010} and the generalized NFW (gNFW) electron pressure profile of
    Ref.~\cite{Planck2013_profile}, we can predict $\langle b_\mathrm{h}P_e \rangle_{\rm HM}$ under different cosmological models and investigate its cosmological dependence.
    Here, we focus on the relationship of $\langle b_\mathrm{h}P_e \rangle_{\rm HM}$ with an ``extended $S_8$ parameter'' defined as
    \ba
    \label{eq:S8prime}
    S'_8 = \sigma_8\left( \frac{\Omega_\mathrm{m}}{0.3} \right)^{\beta_1}\left( \frac{h}{0.7}\right)^{\beta_2}.
    \ea
    The relation is assumed as 
    \ba
    \langle b_\mathrm{h}P_e \rangle_{\rm HM} = A(S'_8)^\alpha.
    \ea   
    We use Eq.~\eqref{eq:hmbP} to calculate $\langle b_\mathrm{h}P_e \rangle_{\rm HM}$ for various values of $\Omega_\mathrm{m}$, $\sigma_8$ and $h$. We find the best-fitting values of $A$, $\alpha$, $\beta_1$, and $\beta_2$ by minimizing $\chi^2 = \left[\langle b_\mathrm{h}P_e \rangle_{\rm HM} - A(S'_8)^\alpha\right]^2$.

    In Fig.~\ref{fig:bPe_S8} we show the relationship between $\langle b_\mathrm{h}P_e \rangle_{\rm HM}$ and $S'_8$ and its best fit at several redshifts. The amplitude of $\langle b_\mathrm{h}P_e \rangle_{\rm HM}$ has a positive correlation with the matter fluctuation as expected.
    We find that $\langle b_\mathrm{h}P_e \rangle_{\rm HM} \propto (\sigma_8\Omega_\mathrm{m}^{0.81} h^{0.67})^{3.14}$ at $z=0$ and that the values of $\alpha$, $\beta_1$, and $\beta_2$ strongly depend on the redshift.  

    We have obtained $\langle b_\mathrm{h}P_e \rangle$ from the halo model, using the gNFW pressure profile of Ref.~\cite{Planck2013_profile}. Therefore, astrophysical effects that impact the total pressure associated with halos would influence the fitting values of $\alpha$, $\beta_1$, and $\beta_2$. 
    
    \subsection{Pressure statistics from observations}
    \begin{figure*}
        \includegraphics[width = \textwidth]{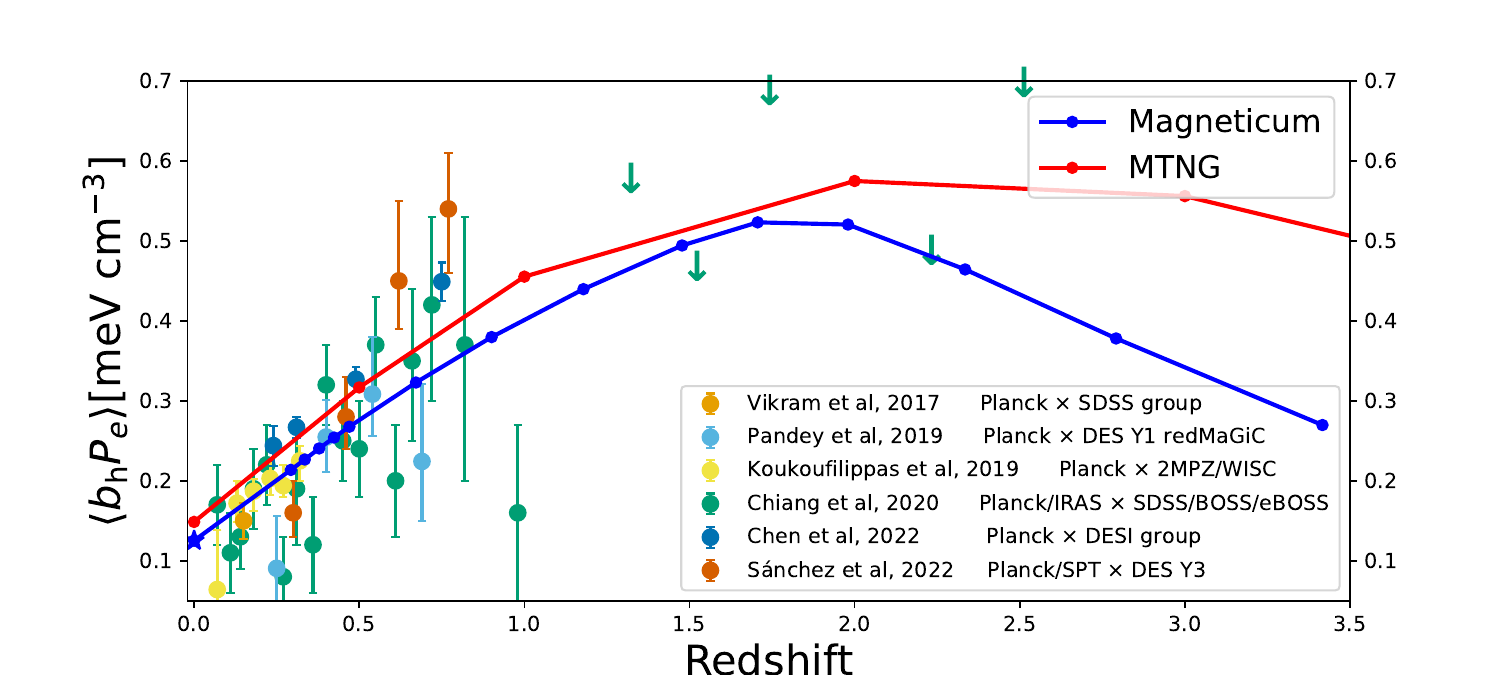}
        \caption{\label{fig:bPe_main_result}Comparison of $\langle b_\mathrm{h}P_e \rangle$ from cross-correlation measurements in observations and simulations. The solid red and blue lines are the MTNG and \emph{Magneticum} simulations, respectively. For the \emph{Magneticum} simulations we choose `Box2b', which has the highest mass resolution. However, `Box2b' lacks data at $z=0$. The data point at $z=0$ is from `Box0' and is marked with a star symbol.   The points with error bars are previous cross-correlation measurements from observational data \cite{Vikram2017, Pandey2019, Koukoufilippas2020, Chiang2020, Chen2022, Sanchez2022}.}
    \end{figure*}
    
    In recent years, several works have measured the large-scale cross-correlation between Compton-$y$ maps and galaxies or galaxy groups. Ref.~\cite{Vikram2017} provided the first such measurement, using the Modified Internal Linear Combination Algorithm (MILCA) Compton-$y$ map from \emph{Planck} \cite{Planck2016_tSZ} and a galaxy group catalog \cite{Yang2007} constructed from the Sloan Digital Sky Survey (SDSS) Date Release 4 (DR4) spectroscopic galaxy survey at a mean redshift of $z=0.15$. Ref.~\cite{Pandey2019} also measured the projected correlation function combining publicly available 2015 \emph{Planck} maps \cite{Planck2016_tSZ} and the redMaGiC catalog from the Dark Energy Survey (DES) Year 1 observation \cite{Flaugher2015}. The redshift range of this measurement is $z=0.15-0.9$ in four bins with a sky coverage of 1321 deg$^2$. Ref.~\cite{Koukoufilippas2020} used the \emph{Planck} 2015 map \cite{Planck2016_tSZ} and a set of photo-$z$ galaxy catalogs, Two Micron All Sky Survey (2MASS) Photometric Redshift catalog \cite{Bilicki2014} and WISE $\times$ SuperCOSMOS catalog \cite{Bilicki2016}. They measured both the angular power spectrum of the galaxies and their cross-correlation with the Compton-$y$ parameter, $C_{\ell,\mathrm{gg}}$ and $C_{\ell,y\mathrm{g}}$. 

    Ref.~\cite{Chiang2020} was the first to constrain $\langle b_\mathrm{h}P_e \rangle$ to a higher redshift ($z>1$). 
    To avoid the leakage of the cosmic infrared background (CIB) in the \emph{Planck} Compton-$y$ maps, they performed correlation measurements separately for the 100, 143, 217, 353, 545, and 857 GHz intensity maps of \emph{Planck} \cite{Planck2016_HFI} and the 100 and 60 $\mu{\rm m}$ maps of the Infrared Astronomical Satellite (IRAS) \cite{IRIS2005} with SDSS, BOSS, and eBOSS \cite{Reid2016, Bautista2018, Schneider2010, Paris2017, Ata2018} in several redshift bins. They compared the results with \textit{Planck} MILCA and Needlet Independent Linear Combination (NILC) $y$ maps and found that the \textit{Planck} $y$ maps show unphysical results above $z=1$. They found that the current CIB model underestimates the CIB intensity at high redshift. Therefore, they only report the upper limit on $\langle b_\mathrm{h}P_e \rangle$ at $z>1$.

    Ref.~\cite{Chen2022} fits the one- and two-halo terms of the $y$-profile simultaneously by stacking the \emph{Planck} 2015 MILCA $y$ map \cite{Planck2016_tSZ} at the position of the DESI galaxy groups \cite{Yang2021}. They divided the group catalog into four redshift bins and divided the mass bins according to the mass distribution of the group. They constrained $\langle b_\mathrm{h}P_e \rangle$ by combining all two-halo term measurements in each redshift bin. Due to the large catalog size and precise mass estimate, the measurements extended to the $10^{13} M_\odot/h$ group mass and had a high signal-to-noise ratio.

    Ref.~\cite{Sanchez2022} uses the magnitude-limited lens sample (MAGLIM) \cite{Porredon2021} from the DES Year 3 observation combining the $y$ maps from the South Pole Telescope (SPT)-SZ + \emph{Planck} \cite{Bleem2022} and \emph{Planck} MILCA \cite{Planck2016_tSZ} for the southern and northern part of the DES footprint, respectively. In large-scale analysis, they use the scale of $k\leq 0.7~ \rm Mpc^{-1}$. The fitting procedure is similar to that of Ref.~\cite{Koukoufilippas2020} but with more Halo Occupation Distribution (HOD) parameters.
    
    In Fig.~\ref{fig:bPe_main_result} we compare the $\langle b_\mathrm{h}P_e \rangle$ measurements from the above literature with those from simulations (Sec.~\ref{sec3}). The results will be discussed in Sec.~\ref{sec4}.

\section{Simulations}\label{sec3}
    To investigate large-scale pressure statistics, cosmological simulations must satisfy two conditions. 
    First, their hydrodynamics needs to accurately model baryonic processes.
    Second, the box size needs to be large enough to contain linear modes and to form at least some rare, massive halos. As discussed in Appendix~\ref{app:poisson_noise}, in a small-volume simulation, the Poisson noise for massive halos causes a nonnegligible effect on the $\langle b_\mathrm{h}P_e \rangle$ measurement.
    
\begin{figure}
        \includegraphics[width = 0.5\textwidth]{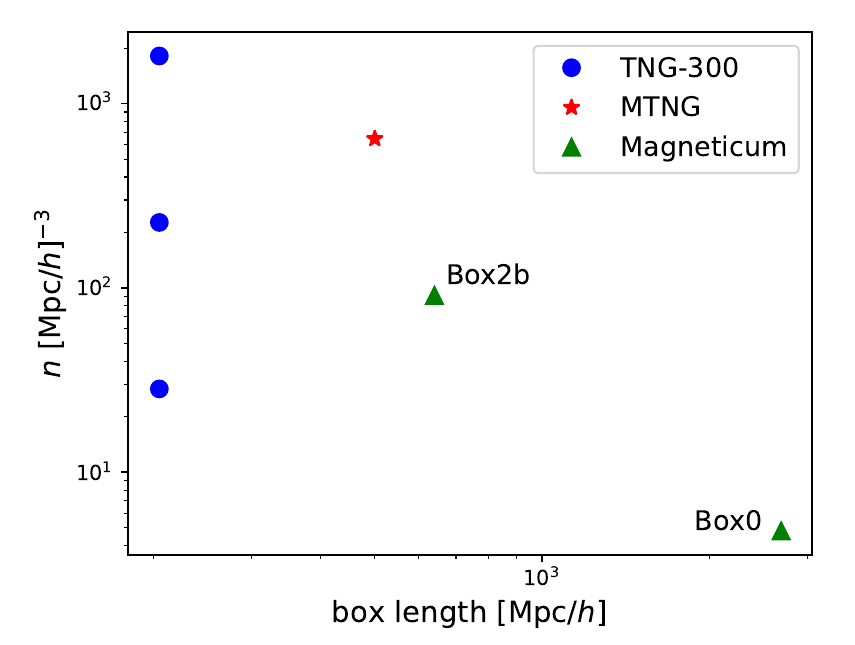}
        \caption{\label{fig:simu_boxlen_resolution} Comparison of the box length and mass resolution of the simulations used in this paper. The $x$-axis gives the box size in units of comoving Mpc/$h$. The $y$-axis indicates the mean number density of dark matter particles. The green triangles represent the \emph{Magneticum} simulations whose names are given next to them. The red star represents the MTNG simulation. The blue circles show TNG300-1, TNG300-2, TNG300-3, from top to bottom.}
    \end{figure}

    In this work, we analyze three simulations: IllustrisTNG (TNG300 \cite{ITNG}), MillenniumTNG (MTNG \cite{Pakmor:2023,Hernandez-Aguayo:2023}) and \emph{Magneticum} \cite{Magneticum}. TNG300 and MTNG have nearly the same galaxy formation model (although their treatments of magnetic fields differ), while \emph{Magneticum} implements a significantly different galaxy formation model. The simulations are described below, and their box sizes and particle numbers are summarized in Table~\ref{tab:simu}. The box length and mass resolution of these simulations are compared in Fig.~\ref{fig:simu_boxlen_resolution}.

    \begin{table*}
    \caption{The basic parameters of the simulations: the simulation's name, the length of the simulation box, the number of dark matter particles or gas cells, the mass of dark matter particles and the mean mass of gas cells. \label{tab:simu}}
    \begin{ruledtabular}
    \begin{tabular}{ccccc}
     Name & Box Size [${\rm Mpc}/h$] & $N_{\rm DM, gas}$ & $M_{\rm DM}~[M_\odot/h]$ & $M_{\rm gas}~[M_\odot/h]$\\
     \hline
	TNG300-1 & 205 &  $2500^3$ & $4.0\times10^7$ & $7.6\times 10^6$ \\
	TNG300-2 & 205 &  $1250^3$ & $3.2\times 10^8$ & $5.9 \times 10^7$\\
	TNG300-3 & 205 & $ 625^3$ & $2.5 \times 10^9$ & $4.8 \times 10^8$\\
	MTNG     & 500 & $4320^3$ & $1.1 \times 10^8$ & $2.1 \times 10^7$\\
	\emph{Magneticum} (Box0) & 2688& $4536^3$ & $1.3 \times 10^{10}$ & $2.6\times 10^9$\\
    \emph{Magneticum} (Box2b) & 640  & $2880^3$ & $6.9\times 10^8$ &$1.4\times 10^8$ \\
    \end{tabular}
    \end{ruledtabular}
    \end{table*}

    \subsection{IllustrisTNG and MTNG}
    MTNG~\cite{Pakmor:2023,Hernandez-Aguayo:2023} is a successor to the IllustrisTNG simulations~\cite{ITNG}, designed for the study of large-scale physical processes \cite{Ferlito2023} and the interpretation of upcoming large-scale galaxy surveys \cite{Contreras2023, Bose2023, Hadzhiyska:2022ugr, Hadzhiyska:2022tvd}. 
    Both simulations utilize the AREPO moving-mesh code \cite{AREPO_code} and a cosmological model based on the findings of Ref.~\citep{Planck2016_cosmo}, with $\Omega_\mathrm{m} = 0.3089$, $\Omega_\mathrm{b} = 0.0486$, $\sigma_8=0.8159$, $n_s=0.9667$, and $h=0.6774$. Their galaxy formation model includes radiative cooling of primordial gas, metal line cooling, an effective model for star formation and the interstellar medium (ISM) \citep{Springel2003}, an effective model for galactic winds, and enrichment from stellar winds and supernovae \citep{Pillepich2018}. It also follows the growth of supermassive black holes and their feedback as AGN \citep{Weinberger2017}. The MTNG model is almost identical to the IllustrisTNG model with only a few minor modifications \cite{Pakmor:2023,MTNG}.
    
    In this study, we focus primarily on the MTNG740 simulation, which employs $4320^3$ dark matter particles and initially $4320^3$ gas cells within a cubic periodic box with a side length of 500 ${\rm Mpc}/h$. 
    To investigate the resolution dependence, we also utilize the TNG300 simulations. These simulations have a side length of 205 ${\rm Mpc}/h$, and were run at three resolutions with a number of dark matter particles/gas cells of $2500^3$, $1250^3$, and $625^3$.
    
    \subsection{\emph{Magneticum}}
    \emph{Magneticum} is a suite of state-of-the-art hydrodynamical simulations based on P-Gadget3 \cite{Springel2005}. Its cosmological model follows the WMAP7 results \cite{Komatsu2011}, with $\Omega_\mathrm{m}=0.272$, a baryonic fraction of 16.8\%, $n_s=0.963$, $\sigma_8 = 0.809$, and $h=0.704$. The baryonic processes include radiative cooling \cite{Wiersma2009}, CMB/UV/X-ray background \cite{Haardt2001}, cooling of 11 elements with the CLOUDY code \cite{Ferland2017},  multiphase ISM \cite{Springel2003}, chemical evolution \cite{Tornatore2007} and AGN feedback \cite{Springel2005_feedback, Fabjan2010}. For more details, see the previous work using this simulation \cite{Dolag2015,Magneticum, Bocquet2016, Gupta2017, Young2021}.

We employ the `Box0' and `Box2b' simulations of \emph{Magneticum}. The `Box0' run has the largest volume with a side length of $2688\ {\rm Mpc}/h$ and the number of particles / cells set to $4536^3$. The `Box2b' run has higher resolution, with $2880^3$ particles/cells in a box of length $640~ \mathrm{Mpc}/h$. 

    Ref.~\cite{Young2021} has investigated the thermal history of the universe by adopting the `Box0' simulation of \emph{Magneticum}. They find that the value of $\langle b_\mathrm{h}P_e \rangle$ from \emph{Magneticum} agrees well with the data reported in Ref.~\cite{Chiang2020}.
    
\subsection{Differences in the simulations}
    \emph{Magneticum} and IllustrisTNG/MTNG employ the same model for ISM and use similar tables for radiative cooling. They use a different UV background, and most importantly a significantly different model for galactic winds and the growth of and feedback from black holes. In addition, IllustrisTNG employs a full treatment of MHD, while MTNG and \emph{Magneticum} do not. \emph{Magneticum} includes an explicit treatment of isotropic thermal conduction with a suppressed Spitzer value of $\kappa=1/20$.

    \subsection{ \texorpdfstring{$\langle b_\mathrm{h}P_e \rangle$}{<bhPe>} in simulations}
    \begin{figure*}
        \includegraphics[width = 0.8\textwidth]{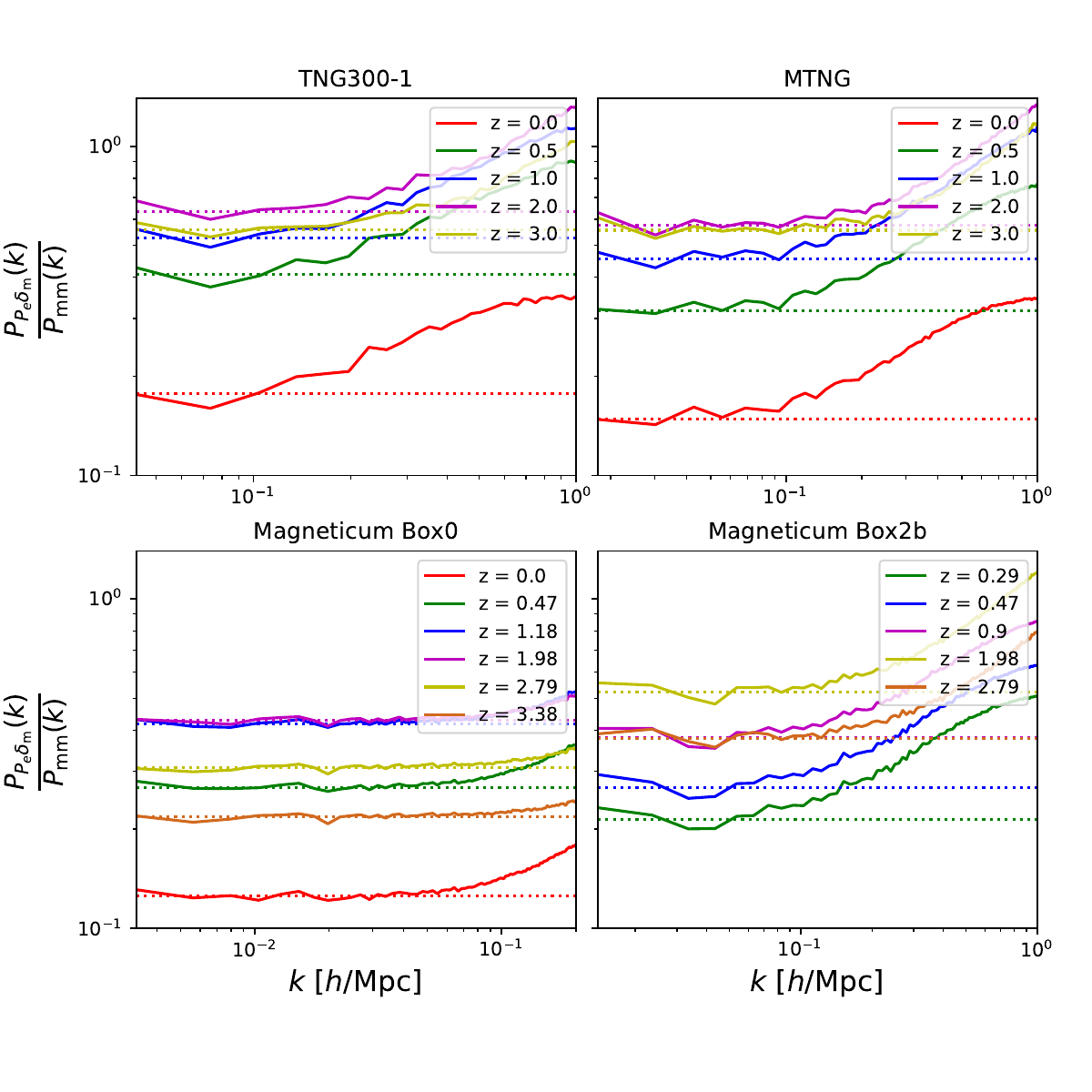}
        \caption{The solid lines represent the ratio of the matter-electron pressure cross power spectrum and the matter auto power spectrum as a function of $k$ in  simulations, TNG300-1, MTNG and \emph{Magneticum} (`Box0' and `Box2b').  Note that the coloring of the redshifts for \emph{Magneticum} is different from the other two. The dotted lines indicate the fitted or average $\langle b_\mathrm{h}P_e \rangle$ values at each redshift. The ranges of $k$ shown in these panels are different due to the different box sizes of the simulations.\label{fig:ratio_Ppem_Pmm}}
    \end{figure*}
    
    To estimate $\langle b_\mathrm{h}P_e \rangle$ in the simulations, we distribute the particles and gas cells on grids of size $512^3$ using the cloud-in-cell assignment scheme. To construct the $\delta_\mathrm{m}$ grid, we add the masses of the particles and gas cells at each grid site. 
    
    To construct the $P_e$ grid, we first sum the internal energy of all gas cells at each grid site. We then convert this to temperature by assuming full ionization and the primordial abundance of hydrogen and helium for the mean molecular mass. Next, we construct the electron number density mesh by summing the masses of gas cells at each grid site and converting this baryonic mass to the electron number density, again assuming full ionization and the primordial abundance. We checked that these assumptions have no effect on the results using the element abundances and ionization fractions stored in the TNG300 runs. We then calculate the electron pressure, $P_e = k_{\mathrm{B}}n_e T_e$, from the electron temperature and number density grids.  

    We Fourier transform the $\delta_\mathrm{m}$ and $P_e$ grids and compute their auto power spectra. We then estimate $\langle b_\mathrm{h}P_e \rangle$ for each simulation snapshot using Eq.~\eqref{eq:bysim} (solid lines in Fig.~\ref{fig:ratio_Ppem_Pmm}). 
    For the \emph{Magneticum} `Box0' simulation, we follow the method in Ref.~\cite{Young2021} by averaging the value of $\langle b_\mathrm{h}P_e \rangle=\langle P_{P_e \mathrm{m}}(k)/P_{\mathrm{mm}}(k)\rangle$ over $k<0.03~ h/{\rm Mpc}$ and ignoring the contribution of the one-halo term. 
    However, for the \emph{Magneticum} `Box2b', MTNG and especially TNG300 simulations, we can only measure the ratio at a larger $k$ due to the smaller box size, where the contribution of the nonlinear component is not negligible. Therefore, we fit the ratio $P_{P_e\mathrm{m}}/P_{\mathrm{mm}}$ as a linear function of $k^2$ at low $k$ (Fig.~\ref{fig:ratio_Ppem_Pmm}),
    \ba
    \frac{P_{P_e\mathrm{m}}}{P_{\mathrm{mm}}}(k) \simeq \langle b_\mathrm{h}P_e \rangle + b_2 k^2 \, ,
    \label{eq:Pmpe_Pmm}
    \ea
    and extract the constant term as our estimate of $\langle b_\mathrm{h}P_e \rangle$.
    In addition, we find that the approximation given in Eq.~\eqref{eq:bysim} does not hold at low redshift. At $z=0$, the one-halo term would be of order $5\%$ of the total cross power spectrum, $P_{P_e \mathrm{m}}(k)$, on a large scale, $k\simeq 0.01~ h/{\rm Mpc}$, according to the halo model. This will be discussed in Appendix \ref{app:1h_term}.

\section{Results}\label{sec4}

    In Fig.~\ref{fig:bPe_main_result}, we compare the mean bias-weighted pressure, $\langle b_\mathrm{h}P_e \rangle$, obtained from the MTNG and \emph{Magneticum} simulations using Eq.~\eqref{eq:Pmpe_Pmm} with observations. In this section, we consider the low ($z<2$) and high ($z>2$) redshift regimes of these results. We then present results from the TNG300 simulations to assess the convergence of $\langle b_\mathrm{h}P_e \rangle$ with respect to the mass resolution of the simulation.
    
    \subsection{Low redshift}
    
    \begin{figure}
        \includegraphics[width = 0.5\textwidth]{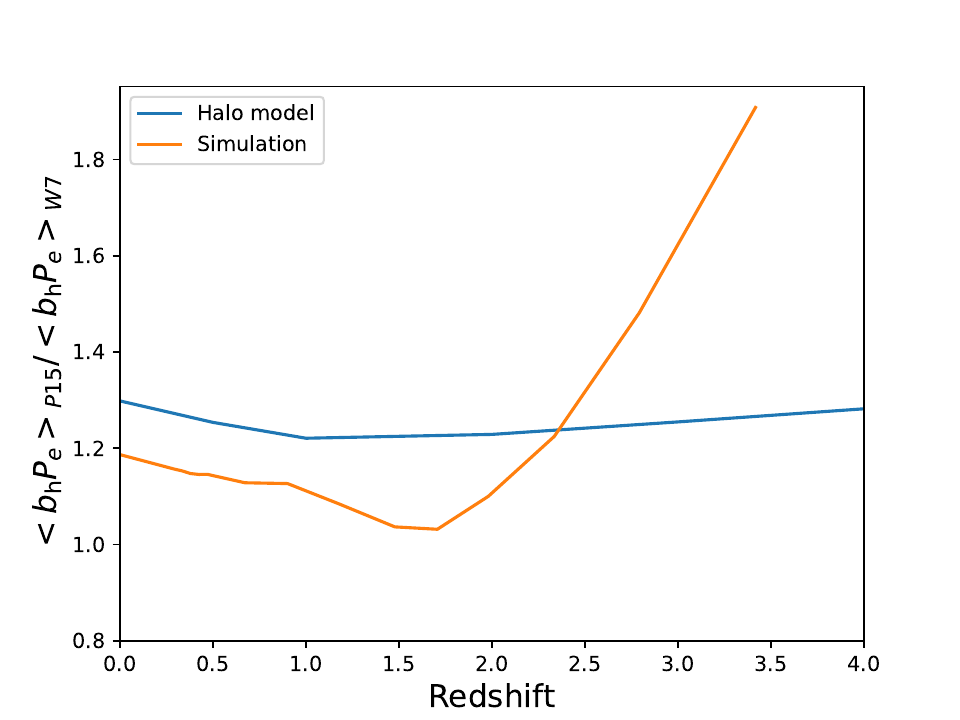}
        \caption{Cosmological parameter dependence of $\langle b_\mathrm{h}P_e \rangle$. The blue line is the ratio of $\langle b_\mathrm{h}P_e \rangle$ predicted by the halo model with \emph{Planck} 2015 and WMAP7 parameters. The orange line is the ratio of $\langle b_\mathrm{h}P_e \rangle$ predicted by MTNG and \emph{Magneticum} simulations (`Box2b'). \label{fig:bPe_diff_cosmo}}
    \end{figure}
 
     At low redshift, the differences between the two simulations are smaller than the current error bars of the observations (Fig.~\ref{fig:bPe_main_result}). This suggests that the difference in their galaxy formation models has little effect on $\langle b_\mathrm{h}P_e \rangle$ below $z=2$.
     
     We find that MTNG predicts systematically larger values for $\langle b_\mathrm{h}P_e \rangle$ by approximately 10\% compared to \emph{Magneticum}. This is expected from the difference in cosmological parameters. The \emph{Planck} 2015 cosmology adopted in MTNG has larger $\Omega_\mathrm{m}$ and $\sigma_8$ than the WMAP7 cosmology adopted in \emph{Magneticum}. The clustering of electron pressure should be higher under \emph{Planck} 2015, as shown in Figs.~\ref{fig:bPe_S8} and \ref{fig:bPe_diff_cosmo}.

     Looking more closely at Fig.~\ref{fig:bPe_diff_cosmo}, we find that the differences between the two simulations are slightly smaller than what we expect based on the halo model calculations. A precise quantification of the expected discrepancy due to the shift in cosmology requires a set of simulations with a fixed galaxy formation model and varied cosmological parameters, which is currently unavailable.
    
    Both simulations predict values for $\langle b_\mathrm{h}P_e \rangle$ that are systematically higher than the observations at $z<1$. By minimizing 
    \begin{equation}
        \chi^2=\sum_{z<z_{\rm max}}\frac{\left(A\langle b_{\rm h}P_e \rangle_{\rm MTNG}-\langle b_{\rm h}P_e \rangle_{\rm obs}\right)^2}{\sigma^2_{\langle b_{\rm h}P_e \rangle_{\rm obs}}}\,,
    \end{equation} 
    we find the best-fitting amplitude, $A$, and its $1\sigma$ uncertainty, which describes how much lower the $\langle b_{\rm h}P_e \rangle$ is in the observations than in the MTNG simulation. We use linear interpolation to find values of $\langle b_{\rm h}P_e \rangle_{\rm MTNG}$ in the redshifts of the observations. We find $A=0.926 \pm 0.024$ for $z_{\rm max}=0.5$. This result is robust, and we find $A\simeq 0.93$ up to $z_\mathrm{max}\simeq 0.75$. The last 2 data points then pull $A$ closer to unity for $z_\mathrm{max}\simeq 1$.
    Although it is still of modest statistical significance, this result is similar to the $S_8$ tension, or the ``lensing is low'' problem \cite{Leauthaud2017} from the lensing observations. 
    
    One caveat to our analysis here is that we have assumed that all data points in Fig.~\ref{fig:bPe_main_result} are independent, which is probably incorrect. Computing the covariance of data points obtained by different authors is difficult and beyond the scope of this paper, but it would be useful to perform this analysis properly as the measurements become more precise.

    It is possible that both galaxy formation models overestimate the clustering of electron pressure. As the two different galaxy formation models roughly agree at low redshift, they would both need to similarly overestimate the electron pressure. It seems more likely that the details of galaxy formation are less important at low redshift, where the pressure is dominated by massive clusters in regions with deep gravitational wells, and the discrepancy between simulations and observations is another manifestation of the $S_8$ tension. 

    As shown in Sec.~\ref{sec2}, $\langle b_\mathrm{h}P_e \rangle$ has a redshift-dependent relationship with the $S_8'$ parameter defined in Eq.~\eqref{eq:S8prime}.
    Therefore, it could be a powerful probe to provide additional constraints on the cosmological parameters. It has a different dependence on cosmological parameters from other probes, such as weak lensing and the tSZ auto power spectrum~\cite{KK:1999,KS:2002,Bolliet:2018,Makiya:2018}, and can help break the degeneracy.

    \subsection{High redshift}
    
    At high redshift, the two simulations differ dramatically. The differences are much larger than expected from the cosmological parameter dependence. Using Eq.~\eqref{eq:hmbP}, we calculate the ratio of $\langle b_\mathrm{h}P_e \rangle$ in two cosmological models, \emph{Planck} 2015 (MTNG) and WMAP7 (\emph{Magneticum}).  The results shown in Fig.~\ref{fig:bPe_diff_cosmo} indicate that the difference due to cosmology is expected to be nearly independent of the redshift. This suggests that the tSZ effect at high redshift ($z>2$) is sensitive to the details of baryonic processes and galaxy formation.

    The galaxy formation model determines the pressure distribution around the halos. In Appendix \ref{app:pe_profile}, however, we show that the difference in the electron pressure profile for the halos with $M>10^{12}~ M_\odot/h$ in these two simulations is not sufficient to explain the difference in $\langle b_\mathrm{h}P_e \rangle$.

\begin{figure}
        \includegraphics[width = 0.5\textwidth]{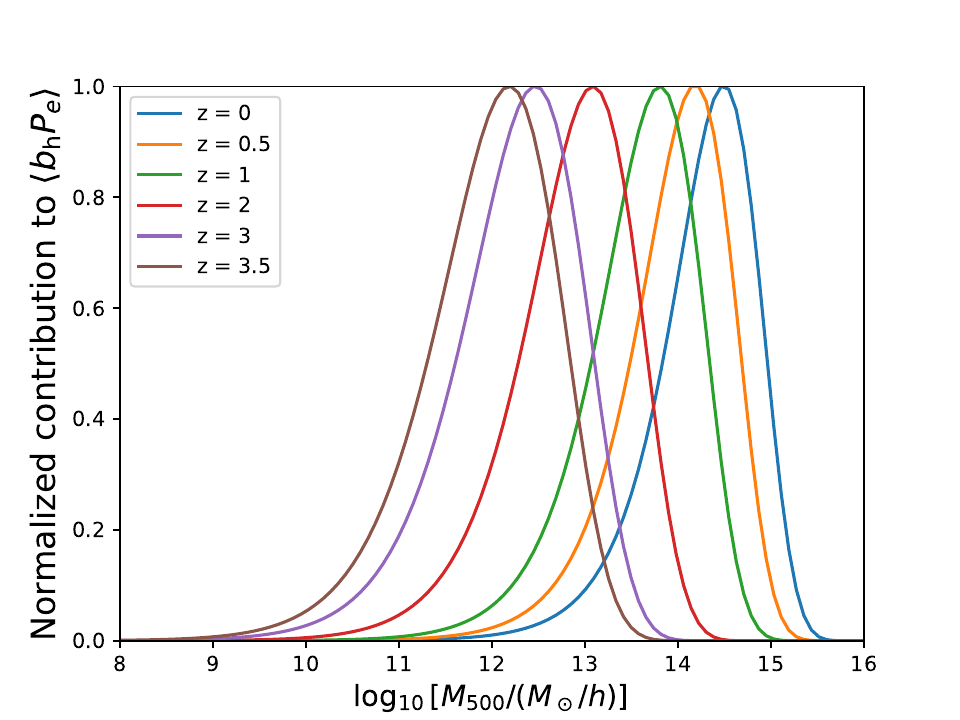}
        \caption{Contribution to $\langle b_\mathrm{h}P_e \rangle$ from different mass ranges at $z=0, 0.5, 1, 2$, and $3$. In the halo model prediction, the contribution from [$M$, $M+{\rm d}M$) is proportional to $P_e^{\rm tot}(M)b_h(M)\dee n(M)$. The peak is normalized to one. \label{fig:contribution}}
    \end{figure}
    
    In Fig.~\ref{fig:contribution}, we show the contribution to $\langle b_\mathrm{h}P_e \rangle$ of each mass bin at different redshift.
    At high redshift ($z>2$), the halos with $M<10^{12}~ M_\odot/h$ become important. These halos, with shallow gravitational potential wells, are more sensitive to baryonic processes, such as stellar and AGN feedback. These processes cannot be well resolved due to poor mass resolution and are determined by subgrid parameters, which are calibrated empirically to broadly match galaxy properties.  

    Both simulations are in modest tension with the upper limits on $\langle b_\mathrm{h}P_e \rangle$ at $z>1$ reported in Ref.~\cite{Chiang2020}. This suggests that at redshifts between $z=1$ and $2.5$, current galaxy formation models predict a higher mean cosmic thermal gas pressure.
    Therefore, future measurements of the large-scale pressure statistic, $\langle b_\mathrm{h}P_e \rangle$, at high redshift will be a valid test for galaxy formation models and will provide guidance for calibrating subgrid parameters. 
    For example, the galaxy formation models in simulations are calibrated by observations, such as the star formation rate density (SFRD), the galaxy stellar mass function (GSMF), and the stellar-to-halo mass relation (SMHM). Except for the SFRD, the other observations (GSMF and SMHM) are from the low-redshift measurements. 
    This may explain the consistency at low redshift between different simulations and may suggest that additional calibrations for galaxy formation models are necessary at high redshift.

    \subsection{Resolution dependence}

    \begin{figure}
    \includegraphics[width = 0.5\textwidth]{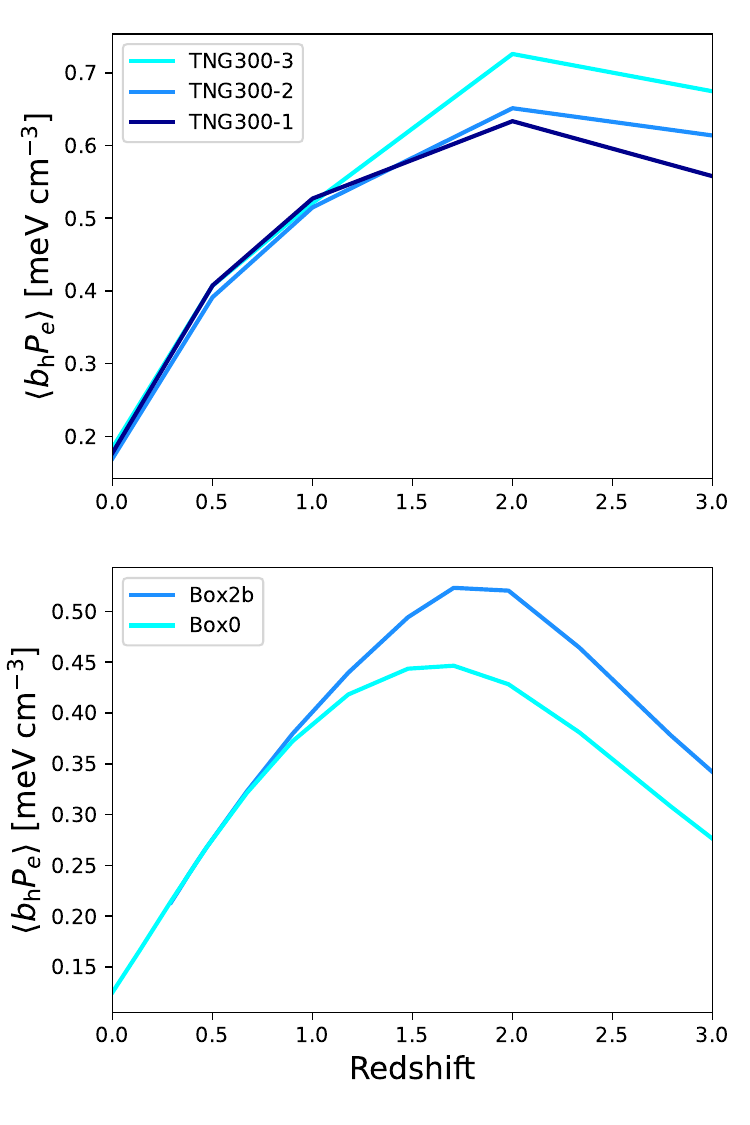}
    \caption{\label{fig:resolution} Comparison of $\langle b_\mathrm{h}P_e \rangle$ as a function of redshift in simulations with different mass resolutions. In the top panel we compare $\langle b_\mathrm{h}P_e \rangle$ derived from TNG300-1 (darkblue), TNG300-2 (lightblue) and TNG300-3 (cyan), whose dark matter particle masses are $4 \times 10^7$, $3.2 \times 10^8$ and $2.5 \times 10^9~ M_\odot/h$, respectively. In the lower panel we compare `Box2b' (lightblue) and `Box0' (cyan) of the \emph{Magneticum}. Their dark matter particle masses are  $6.9 \times 10^8$ and $1.3 \times 10^{10}~ M_\odot/h$, respectively.}
    \end{figure}
    
    In Fig.~\ref{fig:resolution}, we investigate the dependence of $\langle b_\mathrm{h}P_e \rangle$ on mass resolution. The top panel displays comparisons between TNG300-1, TNG300-2, and TNG300-3, with dark matter particle masses of $4.0 \times 10^7$, $3.2 \times 10^8$, and $2.5 \times 10^9~ M_\odot/h$, respectively. 
    At $z > 1$, the amplitude of $\langle b_\mathrm{h}P_e \rangle$ decreases with increasing mass resolution, hinting at a possible convergence with further refinement. The mass resolution of MTNG falls between TNG300-1 and TNG300-2 and is nearly converged.
    The lower panel presents a comparison between the `Box0' and `Box2b' simulations from \emph{Magneticum}, with respective dark-matter particle masses of $1.3 \times 10^{10}$ and $6.9 \times 10^8~ M_\odot/h$. Despite the different volumes of `Box0' and `Box2b', both are sufficiently large to encompass nearly all the massive halos that contribute to the overall $\langle b_\mathrm{h}P_e \rangle$.  Contrary to the TNG300 findings, the \emph{Magneticum} data suggest that the amplitude of $\langle b_\mathrm{h}P_e \rangle$ increases with increasing mass resolution. 

    In the \emph{Magneticum} simulation suite, other available simulations with a mass resolution higher than that of `Box2b' are too limited in box size to accurately measure large-scale pressure statistics, and they lack the large-scale modes necessary to form massive halos, which in turn reduces the amplitude of $\langle b_\mathrm{h}P_e \rangle$. As a result, it is uncertain whether $\langle b_\mathrm{h}P_e \rangle$ would converge with increased mass resolution in \emph{Magneticum} simulations.

    The mass resolution may influence the thermal state of the gas in three manners. First, if the low mass halos could not be resolved, their baryonic processes would be missed. Second, in a lower-resolution simulation, the dark-matter potential wells of halos are more poorly resolved, and these halos would not hold enough baryons to start the baryonic feedback processes. Finally, resolution influences the intensity and start time of some baryonic processes. As discussed in Ref.~\cite{Fabjan2010}, radiative cooling would remove more gas from the hot phase with increasing resolution. On the other hand, higher resolution, which enables a more accurate description, may initiate kinetic feedback from the galactic winds at higher redshift. Also, higher resolution means a lower halo-mass threshold of black hole seeding. The competition between radiative losses and feedback heating decides how resolution influences the thermal state of the gas and its distribution. Consequently, \emph{Magneticum} and TNG300, with their distinct galaxy formation models, exhibit different relationships between mass resolution and $\langle b_\mathrm{h}P_e \rangle$.
    
\section{Discussion}\label{sec5}
    Although MTNG and \emph{Magneticum} use different models of galaxy formation, they give similar results at $z<2$.
    This is because $\langle b_\mathrm{h}P_e \rangle$ at low redshifts is dominated by massive halos ($M > 10^{13}~ M_\odot/h$). Within this mass range, gravitational forces play the main role, and the total pressure associated with a halo is largely determined by its mass. 
    Furthermore, at low redshifts, most baryonic processes tend to be self-regulating.
    To further validate the behavior of $\langle b_\mathrm{h}P_e \rangle$ at low redshifts, we need additional hydrodynamical simulations exploring other galaxy formation models such as BAHAMAS \cite{McCarthy2017}, SIMBA~\cite{SIMBA}, and FLAMINGO~\cite{FLAMINGO}.  

    At $z<1$, there exists a notable discrepancy between the pressure profiles derived from the IllustrisTNG simulations and those measured from the Atacama Cosmology Telescope (ACT) microwave data. The discrepancy goes beyond these simulations, as there is good agreement between the average pressure profiles in the outskirts derived from TNG300, MTNG, \emph{Magneticum}, and the \emph{Planck} and SPT data as shown in Appendix~\ref{app:pe_profile} (also see~\cite{Planck2013_profile}). Ref.~\cite{Amodeo2021} analyzed the stacked radial profile of tSZ up to 6 arcminutes, utilizing a cross-correlation of the CMASS galaxy catalogs and temperature maps derived from the ACT DR5 and \textit{Planck} data. The galaxy sample spans a redshift range of $0.4<z<0.7$ with halo masses of $\simeq 3 \times 10^{13}~M_\odot$, which significantly contributes to the total pressure (Fig.~\ref{fig:contribution}).
    They found that at large radii, the IllustrisTNG simulations substantially underestimate the pressure profile compared to the ACT measurements, with a discrepancy nearly tenfold around 6 arcminutes.

    Ref.~\cite{Moser2023} explored potential systematic uncertainties in both simulations and observations, considering factors such as the line-of-sight integration length, beam smearing effects and angular resolution, halo mass selection, and the two-halo term.
    Most of the uncertainties are from the updated ACT beam, causing around a 10\% variation in the pressure profile, and an uncertain halo selection mass that induces a $\simeq~20\%$ difference. These factors are not sufficient to account for the discrepancy.

    Aligning TNG's pressure profiles with the ACT observations would result in an overestimation of the pressure statistic $\langle b_\mathrm{h}P_e \rangle$, which would conflict significantly with the measurements and imply a total thermal energy in halos exceeding the available gravitational energy \cite{Chiang2021}. 
    More research is needed to elucidate the underlying causes of the discrepancy between the ACT observations and simulations and connect the simulation observables to those in the real Universe \citep{Hadzhiyska2023}.
 
    To measure tSZ at $z>1$, the CIB is an important systematic noise \cite{Chiang2020}. The CIB is a diffuse background radiation in the infrared, mainly due to the cumulative emission from dusty star-forming galaxies throughout the history of the Universe \cite{Chary2001}. Since the star formation rate peaks at redshifts between 1 and 2, the contribution of the CIB during this period is significant. 
    
    Component separation techniques, such as Internal Linear Combination (ILC) \cite{Tegmark2003,Remazeilles2011a,Remazeilles2011b}, used to extract the tSZ signal from the microwave data usually have two assumptions. First, the maps should be a linear combination of components and noise. Second, the components should be uncorrelated. However, the high spatial correlation between CIB and tSZ breaks the second assumption and induces leakage of CIB. This biases the tSZ estimate from ILC methods, as shown in the MILCA/NILC measurements in Ref.~\cite{Chiang2020}. 
    
    To mitigate this contamination, Ref.~\cite{Charmetant2023} uses a constrained ILC (CILC) method, which can cancel more than one contaminant if their spectral energy distributions (SED) are known \cite{Remazeilles2011b}. They find that the CILC method does not significantly reduce CIB noise compared to ILC because the CIB is composed of several spectra and they only use one for deprojection. Worse, additional constraints from the CIB SED reduce the degrees of freedom available for the minimum variance condition, resulting in increased noise for the CILC compared to the ILC. Ref.~\cite{McCarthy2307.01043} attempts to use a moment-based method to reduce the sensitivity to the uncertainties from the CIB SED. Compared to the \emph{Planck} 2015 $y$ map, their method reduces the noise by 10-20\% on the small scale.

    With more precise measurements, $\langle b_\mathrm{h}P_e \rangle$ has the potential to constrain cosmological parameters and shed light on galaxy formation models. To accurately determine the cross-correlation with the tSZ effect with a high signal-to-noise ratio, a comprehensive sample of high-redshift tracers is required. 
    
    High-redshift galaxies and quasars from future deeper galaxy surveys, such as DESI \cite{DESI2022}, PFS \cite{Takada2014}, and the WideField Spectroscopic Telescope (WST) \footnote{https://www.wstelescope.com}, are prime candidates as large-scale tracers. As in Ref.~\cite{RP2023}, the number of quasars identified in the DESI Early Data Release and the first two months of the main survey (EDR+M2) is almost half of the objects in the eBOSS sample. 
    
    The Lyman-$\alpha$ forest is also a potential candidate. The Lyman-$\alpha$ forest refers to a set of absorption lines in the spectra of distant quasars and galaxies caused by the intervening neutral hydrogen gas. With enough samples of quasars, the Lyman-$\alpha$ forest can map the neutral hydrogen clouds in the three-dimensional universe, tracing these large-scale structures \cite{McDonald2003}. 
    
    Ongoing and future CMB experiments such as CCAT-prime~\cite{CCAT-Prime:2023}, Simons Observatory \cite{Ade2019}, LiteBIRD \cite{LiteBIRD2023}, CMB-S4 \cite{Abazajian2016,Abazajian2019}, and CMB-HD \cite{CMB-HD2022} will provide deep and high-resolution CMB measurements. High-frequency channels provided by CCAT-prime can help reduce CIB contamination~\cite{Charmetant2023}. This will allow for a more significant detection of the pressure statistics at $z>1$.

\section{Conclusion}\label{sec6}
    In this paper, we have explored the potential of  tSZ$\times$LSS observables as a probe of cosmology and the galaxy formation model. Specifically, we have focused on the large-scale cross-correlation between galaxies and the tSZ effect. 

    We compared the results of various simulations with observations of the cross-correlation between galaxies and the tSZ effect. We found that the predictions of the \emph{Magneticum} and MTNG simulations are broadly consistent with observations at low redshifts, although there are some discrepancies that may be due to uncertainties in cosmological parameters and the modeling of baryonic physics. 
    
    At high redshift, a large discrepany was found between these two simulations. It is largely due to the modeling of subgrid physics, which is sensitive to mass resolution. The measurements from Ref.~\cite{Chiang2020} indicate a modest tension with the galaxy formation models in both MTNG and \emph{Magneticum} around $z\simeq 2$. It suggests that these galaxy formation models predict a larger $\langle b_\mathrm{h}P_e \rangle$ than observed.
    With more high redshift data from future galaxy survey, better CMB data and more accurate methods to separate CIB comtamintion, the tSZ statistic can help us understand cosmology and galaxy formation better.

\begin{acknowledgments}
We thank G. Efstathiou, K. Osato, and S.~D.~M. White for insightful comments and discussion.
This work was supported in part by the Deutsche Forschungsgemeinschaft (DFG, German Research Foundation) under Germany's Excellence Strategy - EXC-2094 - 390783311. KD acknowledges the support of the COMPLEX project from the European Research Council (ERC) under the European Union’s Horizon 2020 research and innovation program grant agreement ERC-2019-AdG 882679. The calculations for the Magneticum simulations were carried out at the Leibniz Supercomputer Center (LRZ) under the project pr83li and we thank especially for the support by N. Hammer at LRZ when carrying out the Box0 simulation within the Extreme Scale-Out Phase when the SuperMUC Haswell extension system started operation. ZC is supported by the National Science Foundation of China (Nos. 11621303, 11833005, 11890692), National key R\&D Program of China (Grant No. 2020YFC2201602), CSST CMS-CSST-2021-A02, 111 project No. B20019, and the Shanghai Natural Science Foundation, grant Nos. 15ZR1446700 and 19ZR1466800. SB is supported by the UK Research and Innovation (UKRI) Future Leaders Fellowship [grant number MR/V023381/1].
\end{acknowledgments}

\appendix

\section{Halo model}\label{app:halo_model}
    The halo model assumes that all matter can be associated with halos, so the minimal list of ingredients for its construction are the halo mass function, $n_{\mathrm{h}}(M,z)$, the radial density profile of each halo, $\rho_{\mathrm{h}}(r,z|M)$, and the linear halo bias, $b_{\mathrm{h}}(M,z)$. These quantities satisfy the constraints,
    \begin{align}
        \int\dee\log M \frac{\dee n_{\mathrm{h}}(M,z)}{\dee\log M} \frac{M}{\bar\rho_m} = 1 \, , \\
        \int\dee\log M \frac{\dee n_{\mathrm{h}}(M,z)}{\dee\log M} b_{\mathrm{h}}(M,z) \frac{M}{\bar\rho_m} = 1  \, , \\
        4\pi\int_0^{r_\mathrm{out}}\dee r\, r^2 \rho_\mathrm{h}(r,z|M) = M \, .
    \end{align}
    
    As defined in the main text, the matter fluctuations and the electron pressure field can be represented as 
    \ba
        \frac{\rho_{\mathrm{m}}(\mathbf{x}, z)}{\bar{\rho}_{\mathrm{m}}(z)} = \sum_i N_i u_{\mathrm{m,h}}(|\mathbf{x} - \mathbf{x}_i|, z\,|\,M_i) \, ,
    \ea
    and 
    \ba
        \frac{P_{e}(\mathbf{x}, z)}{\bar P_{e}(z)} = \sum_i N_i u_{P_e,\mathrm{h}}(|\mathbf{x} - \mathbf{x}_i|, z\,|\,M_i) \, ,
    \ea
    where $N_i = 0$ or $1$ is the occupation number. Imagine dividing the space into cells that are small enough to contain not more than one halo center. If a cell $i$ contains a halo, $N_i$ is set to 1; otherwise, $N_i$ is set to 0.
    
    The two-point correlation between the matter fluctuation and electron pressure fields is 
    \ba
    \Big< \frac{\rho_{\mathrm{m}}(\mathbf{x}_1, z)}{\bar{\rho}_{\mathrm{m}}(z)} \frac{P_{e}(\mathbf{x}_2, z)}{\bar P_{e}(z)}\Big> & = 
    \Big<\sum_i N_i u_{\mathrm{m,h}}(|\mathbf{x}_1 - \mathbf{x}_i|, z\,|\,M_i) \times \nonumber \\
    & \hspace{-50pt} \sum_j N_j u_{P_e,\mathrm{h}}(|\mathbf{x}_2 - \mathbf{x}_j|, z\,|\,M_j) \Big>\, .
    \ea
    This can be split into a one-halo term ($i=j$) and a two-halo term ($i\neq j$). The one-halo term is 
    \ba
     &\sum_i \langle N_i u_{\mathrm{m,h}}(|\mathbf{x}_1 - \mathbf{x}_i|, z\,|\,M_i)
     u_{P_e,\mathrm{h}}(|\mathbf{x}_2 - \mathbf{x}_i|, z\,|\,M_i)\rangle &
     \nonumber \\ 
     &\hspace{20pt}=
     \int \dee M \frac{\dee n_{\rm h}}{\dee M}\int \dee^3y\ u_{\mathrm{m,h}}(|\mathbf{x}_1 - \mathbf{y}|, z\,|\,M) \times  \nonumber \\
     &\hspace{100pt}  u_{P_e,\mathrm{h}}(|\mathbf{x}_2 - \mathbf{y}|, z\,|\,M)\,.
    \ea
    The two-halo term is 
    \ba
    \sum_i \sum_{j\neq i}\langle N_iN_j u_{\mathrm{m,h}}(|\mathbf{x}_1 - \mathbf{x}_i|, z\,|\,M_i) \times \nonumber \\
    u_{P_e,\mathrm{h}}(|\mathbf{x}_2 - \mathbf{x}_j|, z\,|\,M_j) \rangle
    \nonumber \\
    =1 + \int \dee M_1 \frac{\dee n_{\rm h}(M_1)}{\dee M_1}\int \dee M_2 \frac{\dee n_{\rm h}(M_2)}{\dee M_2} \times \nonumber \\
        \int \dee^3y_1~ u_{m,{\rm h}}(\mathbf{x}_1 - \mathbf{y}_1, z | M_1) \times \nonumber \\
        \int \dee^3y_2~ u_{P_e,{\rm h}}(\mathbf{x}_2 - \mathbf{y}_2, z | M_2) \times \nonumber \\
         b_{\rm h}(M_1)b_{\rm h}(M_2)\xi_{\rm mm}(\mathbf{y}_1 - \mathbf{y}_2)\,,
    \ea
    where $\xi_{\rm mm}(\mathbf{r})$ is the two-point correlation function of matter fields.
    Therefore, the cross correlation function, $\xi_{P_e \mathrm{m}} = \langle \delta_\mathrm{m} P_e\rangle$,  is 
    \ba
        \xi_{P_e \mathrm{m}} = \xi^{1\rm h}_{P_e \mathrm{m}}+\xi^{2 \rm h}_{P_e \mathrm{m}},
    \ea
    with
    \ba
        \xi^{1 \rm h}_{P_e \mathrm{m}} = \bar P_e(z)
        \int \dee \log M \frac{\dee n_{\rm h}}{\dee \log M}  \times  \nonumber \\
         \int \dee ^3y\  u_{\mathrm{m,h}}(|\mathbf{x}_1 - \mathbf{y}|, z\,|\,M)\nonumber \\
      u_{P_e, \mathrm{h}}(|\mathbf{x}_2 - \mathbf{y}|, z\,|\,M)\,,
    \ea
    and
    \ba
        \xi^{2\rm h}_{P_e \mathrm{m}} = & \bar P_e(z)
        \int \dee M_1 \frac{\dee n_{\rm h}(M_1)}{\dee M_1}\int \dee M_2 \frac{\dee n_{\rm h}(M_2)}{\dee M_2} \times \nonumber \\
         &\int \dee^3 y_1~ u_{m,{\rm h}}(\mathbf{x}_1 - \mathbf{y}_1, z | M_1) \times \nonumber \\
         &\int \dee^3 y_2~ u_{P_e,{\rm h}}(\mathbf{x}_2 - \mathbf{y}_2, z | M_2) \times \nonumber \\
         & b_{\rm h}(M_1)b_{\rm h}(M_2)\xi_{\rm mm}(\mathbf{y}_1 - \mathbf{y}_2)\,.
    \ea

    In Fourier space, the profile is transformed as
    \ba
    u(k|M,z) \equiv 4\pi \int^\infty_0 \dee r~ r^2 \frac{\sin(kr)}{kr} u(r|M).
    \ea
    The cross power spectrum, $P_{P_e\mathrm{m}}(k)$, can also be split into one- and two-halo terms 
    \ba
    P_{P_e\mathrm{m}}(k) = P_{P_e\mathrm{m}}^{\rm 1h}(k) + P_{P_e\mathrm{m}}^{\rm 2h}(k).
    \ea
    The one-halo term is 
    \ba
    P_{P_e\mathrm{m}}^{\rm 1h}(k) = \int \dee M \frac{\dee n_\mathrm{h}}{\dee M}  u_{\rm m,h}(k|M,z) \times
    \nonumber \\
    \bar P_e(z) u_{P_e, \rm h}(k|M,z).
    \label{eq:1h}
    \ea    
    The two-halo term is 
    \ba\label{eq:2h}
    P_{P_e\mathrm{m}}^{\rm 2h}(k) &=& 
    P_{\rm mm}(k)\left[ \int \dee M 
     \frac{\dee n_\mathrm{h}}{\dee M} b_{\rm h}(M)
      u_{\rm m,h}(k|M,z) \right] \nonumber \\
     &&\hspace{-20pt}\times
     \left[ \int \dee M 
     \frac{\dee n_\mathrm{h}}{\dee M} b_{\rm h}(M)
      \bar P_e(z) u_{P_e, \rm h}(k|M,z) \right].
      \ea
    Here, the second square bracket,
    \ba\label{eq:HM_bPe}
    \langle b_\mathrm{h}P_e \rangle &=&  \int \dee M 
     \frac{\dee n_{\rm h}}{\dee M} b_{\rm h}(M)
     \bar P_e(z)u_{P_e,\rm h}(k|M,z)  
     \nonumber \\ 
     &&\hspace{-35pt}\xlongequal {k\rightarrow 0} 
     \int \dee M 
     \frac{\dee n_{\rm h}}{\dee M} b_{\rm h}(M) \int^{\infty}_0 \dee r~ 4\pi r^2 P_e(r|M)\,,\nonumber\\
    \ea
    is the bias-weighted mean electron pressure.
    
\section{Poisson noise -- results form TNG300-1}\label{app:poisson_noise}
    \begin{figure}
        \includegraphics[width = 0.5\textwidth]{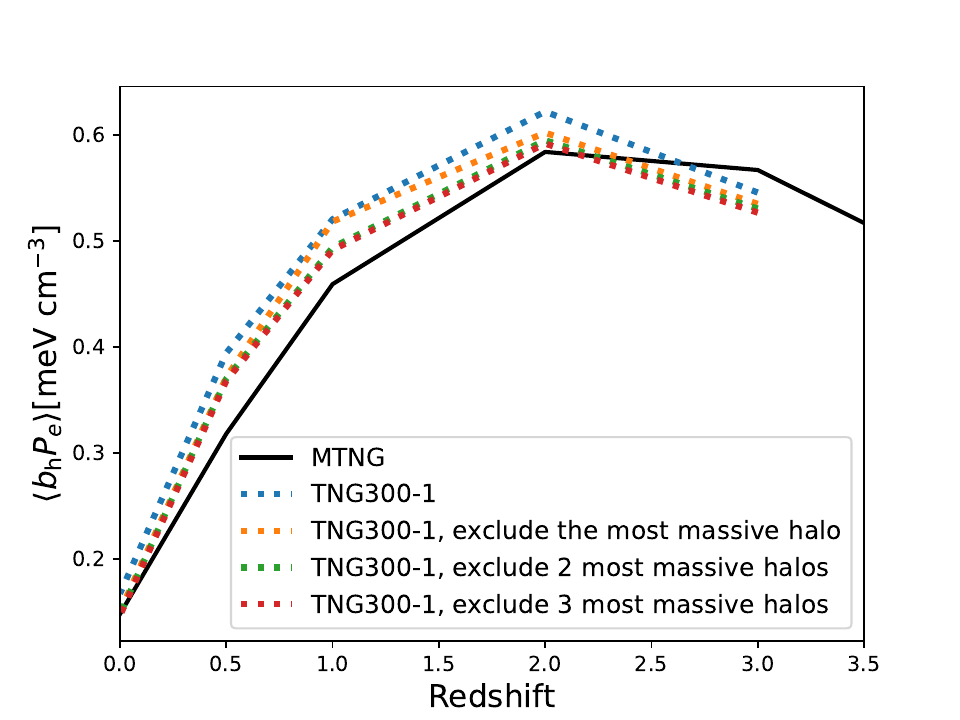}
        \caption{$\langle b_\mathrm{h}P_e \rangle$ from the MTNG (black) and TNG300-1 (blue-dashed) simulations. The orange, green, and red-dashed lines represent masking 1, 2, and 3 of the most massive halos in the TNG300-1 simulation. \label{fig:itng}}
    \end{figure}
    
    The size of the simulation box can influence $\langle b_\mathrm{h}P_e \rangle$ in two ways. First, the lack of large-scale modes in small-box simulations causes a lack of most massive halos. Second, due to the Poisson noise for massive halos, if the size of the simulation box is not large enough, the uncertainty of the massive halo number will be large.
    For example, an anomalous massive halo is found in TNG300-1, making its $\langle b_\mathrm{h}P_e \rangle$ larger than expected. 
    
    In Fig.~\ref{fig:itng}, we compare TNG300-1 and MTNG, masking the most massive halos in TNG300-1. We find that an anomalous massive halo significantly influences the value of $\langle b_\mathrm{h}P_e \rangle$. When we mask this massive halo, the results of TNG300-1 become more consistent with MTNG. 
    
    Note that the cosmic variance and the small modification in the galaxy formation model could also contribute to the difference between TNG300 and MTNG.
    
\section{One-halo term}\label{app:1h_term}

\begin{figure}
        \includegraphics[width = 0.49\textwidth]{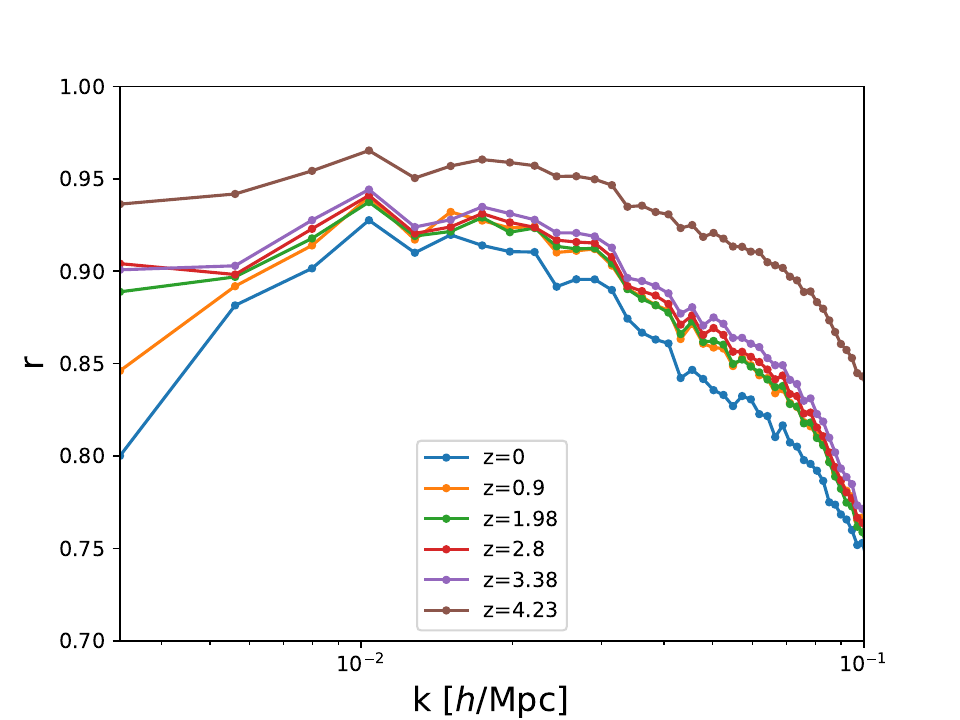}
        \caption{The correlation parameter $r$ [Eq.~\eqref{Eq:r}] from the \emph{Magneticum} `Box0' simulation. \label{fig:r}}
    \end{figure}

    We define the correlation parameter between the electron pressure and matter as
    \ba\label{Eq:r}
        r \equiv \frac{P_{P_e\mathrm{m}}(k)}{\sqrt{P_{\rm mm}(k)P_{P_eP_e}(k)}}.
    \ea
    On large scales, where the density fluctuation is in the linear regime, 
    it is expected that $r$ increases with scale (decreases with $k$) and approaches unity as $k\rightarrow 0$.
    However, in the \emph{Magneticum} simulation (`Box0’), we observe that $r$ increases with $k$, when $k<0.01~ h/{\rm Mpc}$ (Fig.~\ref{fig:r}).
    The \emph{Magneticum} `Box0’ is the only one with modes at $k<0.01~ h/{\rm Mpc}$. In the MTNG simulation, whose box size is not large enough, we do not observe $r$ increasing with $k$.
    
    To address this discrepancy, we explore the matter-electron pressure cross power spectrum using the halo model. 
    To calculate the one- and two-halo terms of the power spectrum, we use Eqs.~\eqref{eq:1h} and \eqref{eq:2h} with the halo mass function of Ref.~\cite{Tinker2008}, the halo bias of Ref.~\cite{Tinker2010}, and the pressure profile of Ref.~\cite{Planck2013_profile}.
    
    We find that, unlike for matter, the one-halo term of the cross power spectrum is not entirely negligible on large scales ($k<0.01~ h/{\rm Mpc}$), as the pressure profile is much more extended than that of matter.
     At $z=0$, the one-halo term constitutes approximately 5\% of the total cross power spectrum. We also find that the one-halo term contribution diminishes and becomes negligible at higher redshifts  ($z>3$). 
    This might elucidate the unexpected behavior of $r$.
    
    However, as highlighted in Refs.~\cite{Schmidt2016, Chen2020}, the large-scale behavior of the one-halo term requires careful consideration. It is unphysical for the one-halo term to remain constant for $k<0.01~ h/{\rm Mpc}$. Otherwise, it would surpass the two-halo term as $k\to 0$. We leave the comprehensive treatment of the large-scale one-halo term and a deeper exploration into the behavior of $r$ to future research.

\section{Pressure profile}\label{app:pe_profile}
    \begin{figure*}
        \includegraphics[width = 1\textwidth]{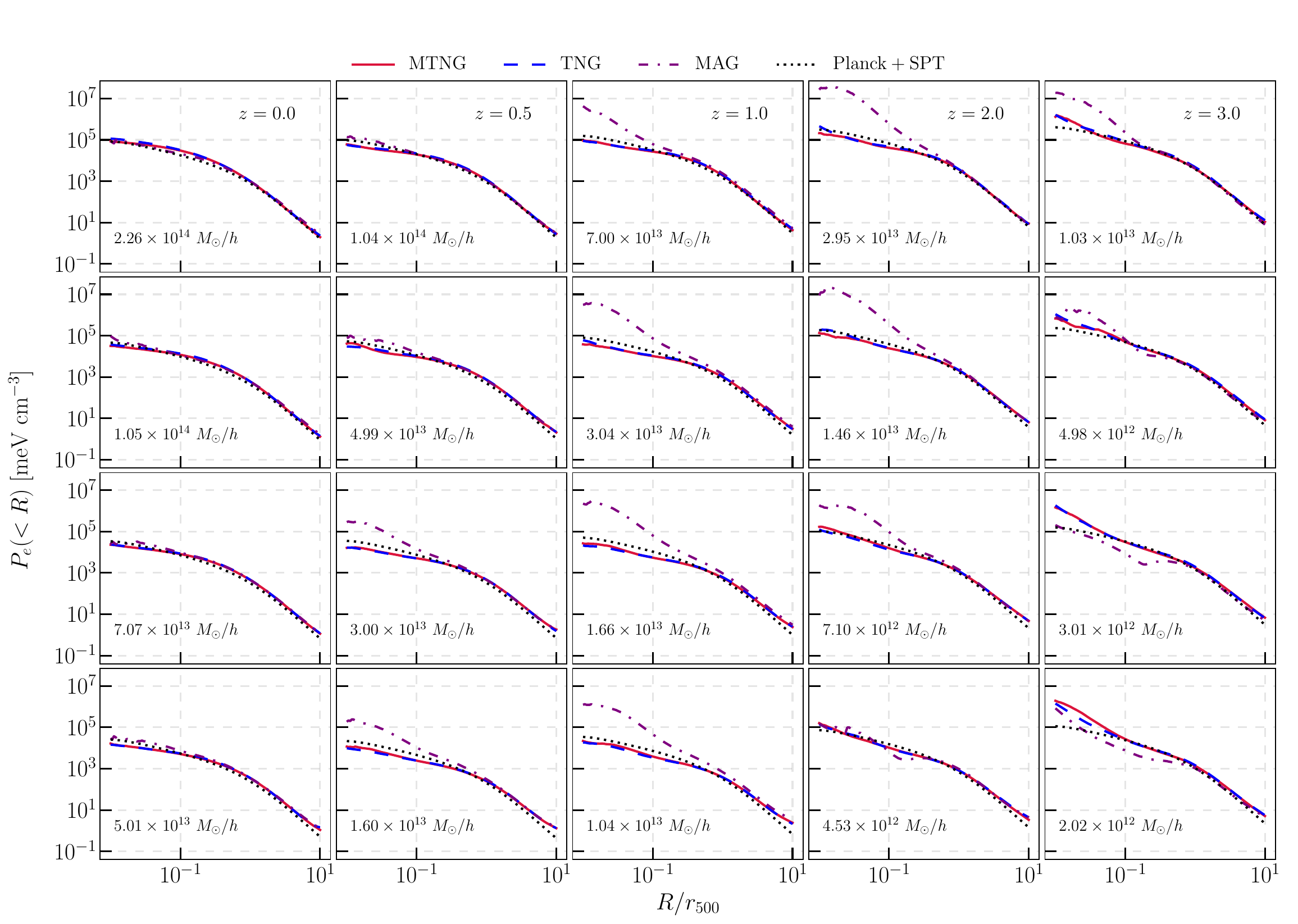}
        \caption{The average pressure profiles within a radius $R$ [Eq.~\eqref{eq:cum}] of MTNG (red), TNG300-1 (blue dashed), \emph{Magneticum} `Box0' (purple dot-dashed), and the gNFW fitting formula of Ref.~\cite{Melin2023} (dotted).\label{fig:profile}}
    \end{figure*}
    
    In this Appendix, we compare the pressure profiles of halos with $M>10^{12}~ M_\odot/h$ in the MTNG, TNG300-1, and \emph{Magneticum} `Box0' simulations. We divide the halos into four mass bins at $z=0$, $0.5$, $1$, $2$, and $3$, and stack the electron pressure profiles of each mass bin.

    In Fig.~\ref{fig:profile} we show the average pressure profiles within a radius $R$, defined as
    \ba
    \label{eq:cum}
    P_e(<R) = \frac 1{V(R)} \int_0^R  P_e(r) \, \dee^3 r,
    \ea
    where $V(R)=4\pi R^3/3$.
    We also show the comparison with the gNFW fitting formula given in Ref.~\cite{Melin2023}.
    The difference between the electron pressure profiles in these two simulations is only apparent in the inner part of the halos, $R<0.2 \,r_{500}$. As shown in Eq.~\eqref{eq:HM_bPe}, the bias-weighted mean electron pressure is determined by the total pressure associated with each halo, rather than by the details of the shape of the pressure profile.
    Therefore, the difference in profile caused by the galaxy formation model is not enough to explain the difference in $\langle b_\mathrm{h}P_e \rangle$ at $z>1$.
    
\bibliographystyle{apsrev4-2} 
\bibliography{references_inspire}
\end{document}